# Micro- Mechanical assembly of high-quality Fabry-Perot microcavities for the integration with two-dimensional materials

Christoph Rupprecht[1], Nils Lundt[1], Sven Höfling[1,2], Christian Schneider[1,3]

[1]*Technische Physik and Wilhelm-Conrad-Röntgen-Research Center for Complex Material Systems, Universität Würzburg, D-97074 Würzburg, Am Hubland, Germany*

[2]*SUPA, School of Physics and Astronomy, University of St. Andrews, St. Andrews KY 16 9SS, United Kingdom*

[3] *Institute of Physics, University of Oldenburg, 26129 Oldenburg, Germany*

*Abstract:*

*Integrating monolayers of two-dimensional semiconductors in planar, and potentially microstructured microcavities is challenging because of the few, available approaches to overgrow the monolayers without damaging them. Some strategies have been developed, but they either rely on complicated experimental settings, expensive technologies or compromise the available quality factors. As a result, high quality Fabry-Perot microcavities are not widely available to the community focusing on light-matter coupling with atomically thin materials.*
*Here, we provide details on a recently developed technique to micro-mechanically assemble Fabry-Perot Microcavities. Our approach does not rely on difficult or expensive technologies, and yields device characteristics marking the state of the art in cavities with integrated atomically thin semiconductors.*

*Introduction:*

Atomically thin transition metal dichalcogenide crystals (TMDCs) belong to an emergent class of materials relevant to studies in fundamental- as well as application-oriented light matter interaction. They combine huge oscillator strength and thus optical activity and giant exciton binding energies, making them a particularly interesting platform for cavity quantum electrodynamics. To date, the regime of strong light matter coupling in optical microcavities has been convincingly demonstrated with single monolayers[1], even at room temperature[2–4]. In order to integrate a TMDC monolayer into a (planar) Fabry-Perot microcavity, a variety of approaches have now been introduced. Typically, at first, a monolayer is placed on a bottom DBR. Subsequently, the monolayer must be capped with some type of dielectric material and finally a top mirror is deposited. This can be a metallic layer or DBR. The deposition of metallic layers has been pursuit by various groups for initial demonstrations of strong light-matter coupling with TMDC monolayers since it is rather straightforward to implement and yields optical resonators with relatively small effective cavity lengths on the order of a few 100 nm [1] . Yet, it faces intrinsic limitations in available Q-factors (typically Q<1000), and thus is prohibitive for a wide range of experiments relying on extended photon lifetimes. Deposition of dielectrics on TMDC monolayers is very challenging since most techniques have a detrimental impact on the luminescence properties of the monolayer. Such deposition was tested with atomic layer deposition (ALD)[5], physical vapor deposition (PVD/sputtering), electron beam evaporation and plasma-enhanced chemical vapor deposition (PE-CVD)[4]. In their default setting, all these techniques damage the monolayer during the deposition process due to a high process temperature, the impact of high-energy ions or a reactive process atmosphere: The photoluminescence after such deposition monolayers is either completely quenched or strongly decreased and broadened as in the case of ALD deposition of $Al_2O_3$.

Yet, there are techniques, which only weakly affect the monolayer properties, and which were successfully tested in microcavity implementations: These are spin-coating of poly-methyl-methacrylate (PMMA)[3] and plasma-assisted evaporation (PAE) of $SiO_2$ [6]. Both techniques are also compatible with TMDC monolayers encapsulated by hexagonal Boron Nitrite.
The successful fabrication of high quality TMDC- microcavities based on PAE has been addressed in recent works [6,7]. However, the vast majority of research groups do not have access to a state of the art PAE system for infrastructural or budget-related reasons.

An alternative approach pursues the idea of utilizing DBR mirrors, which are initially grown on separate substrates. Such DBRs can be commercially purchased by various companies and have been optimized to yield extremely high reflectivity. Our idea then foresees that the top DBR is simply placed mechanically on top of the bottom DBR with the coated monolayer. While the technique itself has been applied by us in a previous work to assemble a full cavity [8], here we provide much more profound insight into its methodology.

*Experiment:*

To implement this idea, a TMDC monolayer is transferred on a $SiO_2/TiO_2$ bottom DBR (10 pairs) that terminates with the low refractive index material. This bottom DBR is purchased from a commercial vendor with 129nm (79 nm) $SiO_2$ ($TiO_2$) layers designed to have the stop band centered at 750nm. We note, that the typical price range for such a DBR mirror is ~ 100 €. Next, the atomically thin crystal is isolated and transferred on the DBR with the conventional PDMS stamping method and a home-built transfer stage. This stamping is well documented in the literature [9], and will not be discussed in greater detail here. To obtain a full cavity including the second half of the cavity corea and the top DBR, PMMA was spin coated for 60s with 6000 turns/sec to approximately match the $\lambda/2$-condition for the resonance wavelength. The PMMA concentration in its solvent Anisole needs to be adjusted in order to obtain the correct cavity thickness. The coated PMMA is baked at a hotplate for 2 minutes at 165 °C. Importantly, PMMA capping is known not to affect the quality of the monolayer- and has been successfully tested with fully encapsulated TMDC monolayers [3]. In stark contrast to non-invasive dielectric capping strategies, it does not rely on expensive technology equipment.
Next, we put the prepared DBR + monolayer aside and take a separate DBR with 8.5 pairs and high refractive index termination. It should be noted, that in our study, the DBR pair numbers are asymmetric, to couple light preferentially towards the top direction, where the signal is detected. Then, we utilize a sharp item, (in our case, a small screwdriver), to manually apply moderate pressure and scratches to the surface of the second $SiO_2/TiO_2$ DBR. This procedure yields fragmentation of the DBR layer coating, and small pieces (10-50 μm diameter) of this DBR dis-attach from the substrate. Since the second DBR is terminated on both sides with $TiO_2$, flipping of the mirror by 180° still results in the desired cavity.

The dis-attached pieces can be picked up with a PDMS gel stamp and can subsequently be transferred on the prepared bottom DBR with the identical transfer method utilized for the TMDC monolayers. Importantly, only pieces with an apparently homogenous and flat surface under inspection with an optical microscope were transferred to avoid strongly shifting the cavity mode by e.g. dirt particles on the surface of the second DBR (see e.g. Fig. S1). If there is dirt of any small particle, the DBR fragments are not transferred. The van-der-Waals force is sufficient to hold both DBRs together. We do not utilize a heated stage, or accurately calibrated pressure in the dry stamping process. This process is illustrated in Fig. 1a) and a microscope image of a full structure is shown in Fig. 1b) with the arrow pointing towards the transferred DBR. Fig. S1 shows microscope images of dis-attached DBR pieces on PDMS and of the same transferred on the PMMA cavity layer

The dispersion relation, measured via angle-resolved reflectivity, is shown in Figure 1b). By calibrating our setup, we can convert the angle of detected light to the in-plane wavevector. Details on the procedure are given, e.g. in [10]. Our experiment reveals a well-pronounced cavity dispersion relation, characterized by an effective mass of $1.23 \times 10^{-5}$ $m_e$ with $m_e$ being the free electron mass.

The line spectrum at , depicted in 1c), shows an extremely narrow cavity mode with a linewidth of 0.163 nm. This corresponds to a Q-factor of $3827 \pm 222$. According to the transfer matrix calculation of this structure, the Q-factor of 25000 can be theoretically achieved, however it is presumably limited by material inhomogeneities (photonic disorder). The calculation also yields an effective cavity length of 400 nm. We fabricated 9 cavities with Q-factors between 500 and 4600.

The PMMA layer on top of the bottom DBR acts as spacer between the DBRs. Its thickness can be varied to adjust the cavity resonance. In order to check the reproducibility, we fabricated a variety of microcavities and reference samples (PMMA layer on $SiO_2$) with different PMMA concentrations: First, we check the physical layer thicknesses of the samples fabricated with different PMMA concentrations using a *Dektak 3030* Profilometer. In order to get a defined step, we use a razor blade to remove a small area of PMMA. We conduct this measurement on various positions of our reference samples, and calculate the average thickness and standard variation. The results are plotted in Fig 2a). We notice a very accurate correlation between PMMA concentration and layer thickness, which yields an almost linear dependence. Next, we fabricate assembled cavities with the same PMMA concentrations, and study the cavity resonance energy. Fig 2b depicts the corresponding cavity energy, which were resolved by white light spectroscopy (red dots).

Here, we observe two classes of devices: The first class of devices follows the expected linear correspondence between cavity thickness d (respectively, PMMA concentration) and optical mode in a Fabry-Pérot cavity[12]. The red line, providing a guide to the eye, visualizes this. The second class of devices display a strong deviation from this behavior (arrows).

To provide an understanding of this peculiarity, we conducted transfer matrix simulations of the nominal device structures. As input parameters, we use the physically extracted PMMA thickness (Fig 2a), the $SiO_2$ and $TiO_2$ thicknesses provided by the vendor of the DBRs, the refractive index data of $SiO_2$ and $TiO_2$ provided by the vendor, and the refractive index of PMMA.[11] The refractive index data was assumed to be without uncertainties, so the plotted ones result from the thickness measurement of PMMA. All measured resonance energies can be rather accurately reproduced by the simulations: The ones following the approx. linear trend between concentration and energy evolve for devices where the top DBR starts on a $TiO_2$ layer. The seemingly deviating cavity resonance for PMMA concentrations of 3.5 % is a higher longitudinal mode. The cavities displaying a strongly deviating resonance energy for concentrations 2.5% can be modelled by devices where the top DBR starts on a $SiO_2$ layer, indicating an intrinsic inaccuracy during breaking the top DBR with the screw driver (here, we conclude that there is a remaining possibility of the mirror breaking at the $TiO_2/SiO_2$ interface). No difference was obtained when comparing the DBRs of these two device classes under an optical microscope (see Fig. S1b,d as example).

Last, we check the polarization properties of our mechanically assembled cavities. We exemplarily study the angle dependent polarization splitting (frequently referred to as TE/TM splitting), evolving from the angle dependent phase delay of light reflected at a dielectric mirror [11]. This polarization phenomena, which is strongly depending on the angle incident and reflected light, is a core resource in advanced experiments dedicated to the study of topological phenomena and spin-orbit coupling of light in microcavities [8,13,14 15] . Fig. 3a) displays a reflectivity spectrum of a cavity with 2.5% PMMA concentration, which is clearly characterized by two optical resonances with a degenerate energy at . As expected, the splitting scales for small incident angles $\theta$ approx. with $\theta^2$ and therefore with $k^2$ ,where $k$ is the in plane wave vector [12] (Fig 3b). The magnitude of the TE/TM splitting depends on a variety of factors, but most notably on the deviation of the cavity energy (determined by thickness and refractive index of the cavity) from the central frequency of the DBR stop band. [12] This behavior is well reflected in Fig. 3c), where we plot the magnitude of the TE/TM splitting extracted at as a function of the cavity mode energy. Indeed, for a cavity energy close to the Bragg condition of the DBR mirrors, the TE/TM splitting approaches 0 meV, while its value approaches giant numbers ~ 20 meV for the case of significant deviations from the Bragg condition.

**Conclusion**

In summary, our paper introduces a low-cost approach to the fabrication of high quality factor microcavities based on micromechanical assembly. Our technique is highly inspired by the dry PDMS transfer method for TMDC fabrication. We utilize commercially available DBR mirrors in the price sector ~ 100 € per piece and a home-built transfer microscope as technology tools. We believe that this technique can be used by small and medium size teams without access to difficult and expensive coating infrastructure to explore the field of cavity quantum electrodynamics with atomically thin crystals.

**Acknowledgement**

We acknowledge support by the state of Bavaria. C.S. acknowledges support by the European Research Commission (Project unLiMIt-2D). Useful discussions with C. Anton-Solanas are acknowledged.

**Figures**

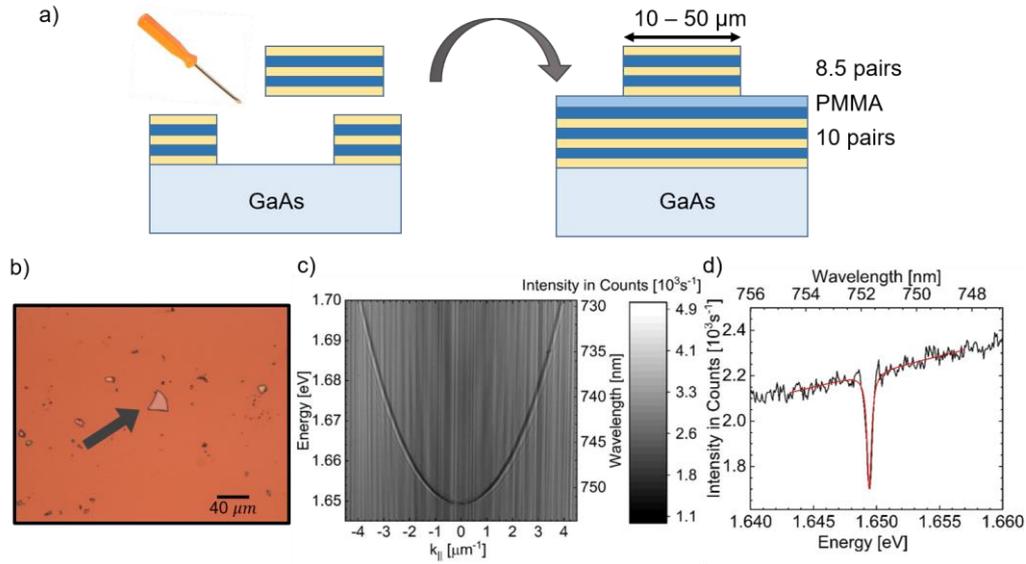

**Fig. 1** a) schematic illustration of the flip chip process. b) example of an exfoliated flip chip mirror. c) high Q photonic energy dispersion plotted against in plane wave vector $k_{\parallel}$ d) Spectrum at $k = 0 \ \mu m^{-1}$ from c) with Lorentzian fit. Extracted $Q = 3827 \pm 222$.

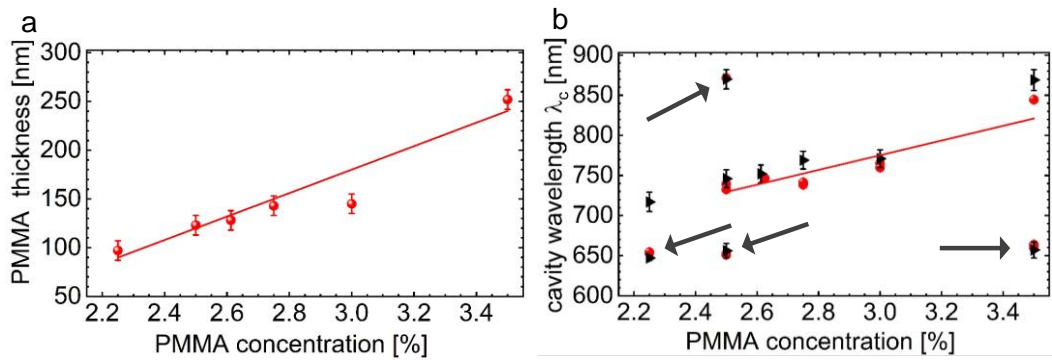

**Fig. 2** a) Measured PMMA thickness with profilometer Dektak 3030 on $SiO_2$ reference samples with guideline for the eye (red). b) measured optical cavity wavelength (red) and transfer matrix simulations (black). Input parameters are the measured thicknesses from a), the refractive index data of PMMA[10] as well as of $TiO_2$ and $SiO_2$ from the vendor. The layer thicknesses of $SiO_2$ and $TiO_2$ are 129nm and 79nm resulting in a DBR stop band center at 750nm. The red line corresponds to guideline for the eye and the black arrows indicate data points deviating from the linear behavior.

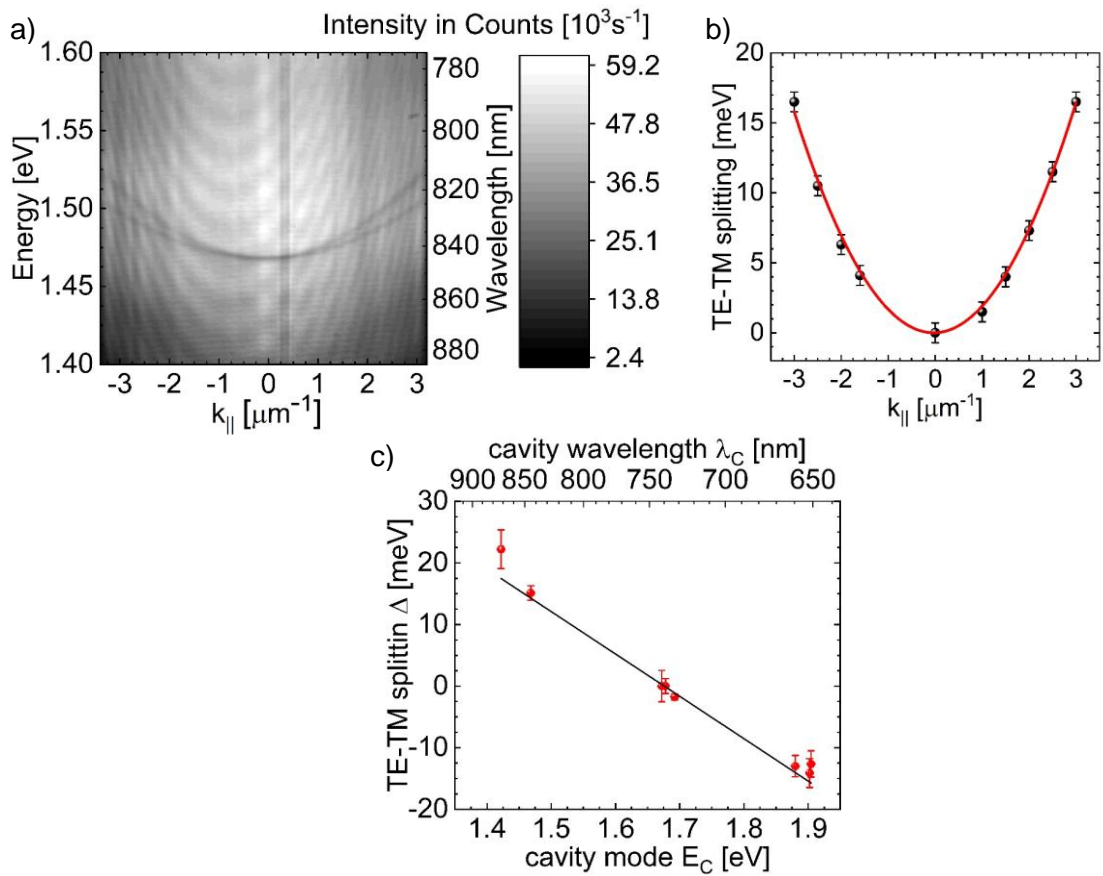

**Fig. 3** a) Energy-momentum dispersion of flip chip cavity with 2.5% PMMA concentration and high TE-TM splitting. b) TE-TM splitting extracted from a) and parabolic fit (red). c) TE-TM splitting at $k_{\parallel} = (3.00 \pm 0.05)$ µm$^{-1}$ for cavities with different PMMA concentrations, hence cavity wavelengths. The black line indicates a guideline for the eyes.

# Supplementary Material: Micro- Mechanical assembly of high-quality Fabry-Perot microcavities for the integration with two-dimensional materials

Christoph Rupprecht[1], Nils Lundt[1], Sven Höfling[1,2], Christian Schneider[1,3]

[1]Technische Physik and Wilhelm-Conrad-Röntgen-Research Center for Complex Material Systems, Universität Würzburg, D-97074 Würzburg, Am Hubland, Germany

[2]SUPA, School of Physics and Astronomy, University of St. Andrews, St. Andrews KY 16 9SS, United Kingdom

[3] Institute of Physics, University of Oldenburg, 26129 Oldenburg, Germany

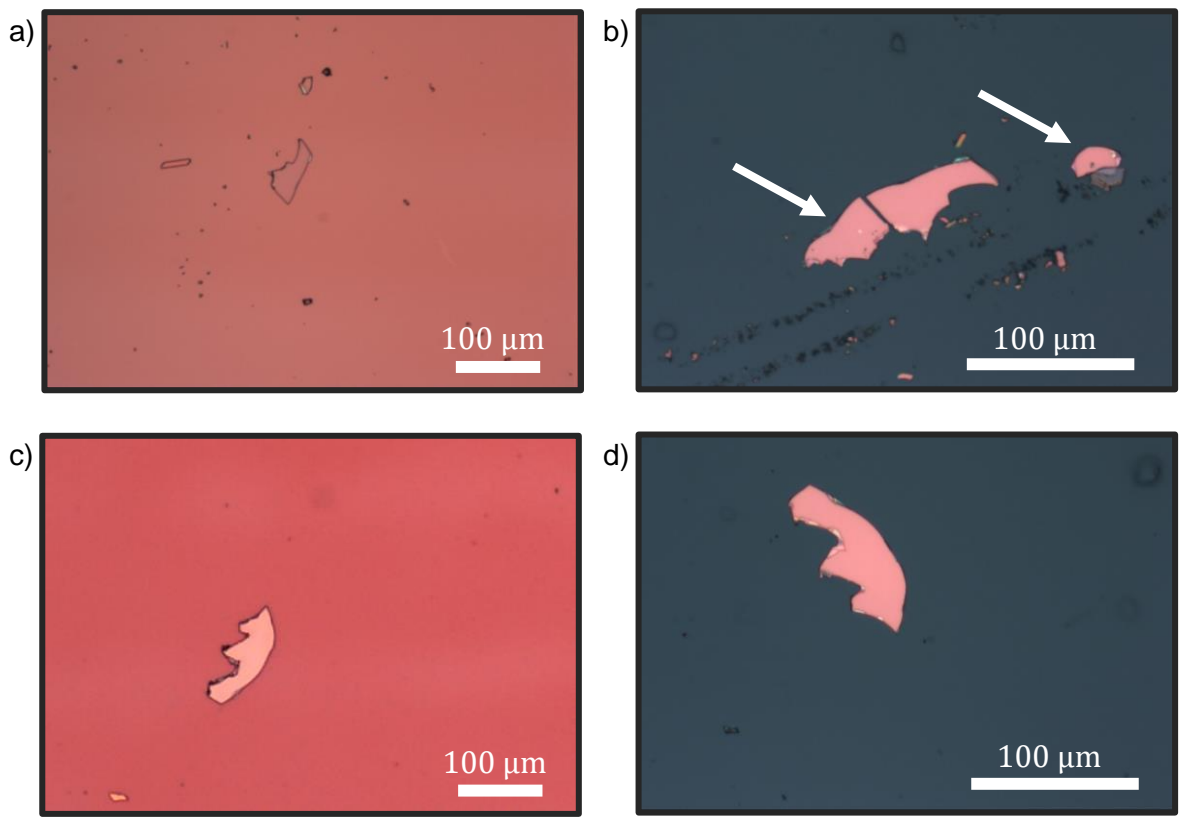

**Fig. S1** a) Transferred top DBR on PMMA with 2.25% concentration. b) DBR on PDMS before transfer . This DBR is terminated with $SiO_2$ according to transfer matrix simulations. White arrows indicate pieces with inhomogeneous surfaces, which were not used to avoid large shifts in the cavity resonance. c) Transferred top DBR on PMMA with 2.6125% concentration. d) DBR on PDMS before transfer. This DBR is terminated with $TiO_2$ according to transfer matrix simulations.